\documentclass[aps,prd,preprintnumbers,floats,nofootinbib,amssymb]{revtex4}
\pdfoutput=1
\usepackage{slashed}
\usepackage{graphicx}
\usepackage{subfigure}
\usepackage{dcolumn}
\usepackage{bm}
\usepackage{color}
\usepackage{ulem}
\usepackage{multirow}
\usepackage{enumitem}
\usepackage{url}
\usepackage{hyperref}
\usepackage{cleveref}
\begin{document}

\baselineskip=15pt


\title{CP violating phase sum rule $\delta^q_{\rm KM}+\delta^l_{\rm KM}=0$ for CKM and PMNS matrices }

\author{Junxing Pan${}^{1}$\footnote{panjunxing2007@163.com}}
\author{Jin Sun${}^{2}$\footnote{019072910096@sjtu.edu.cn}}
\author{Xiao-Dong Ma${}^{3}$\footnote{maxid@phys.ntu.edu.tw}}
\author{Xiao-Gang He${}^{3,4}$\footnote{hexg@phys.ntu.edu.tw}}

\affiliation{${}^{1}$School of Physics and Information Engineering, Shanxi Normal University, Linfen 041004, China}
\affiliation{${}^{2}$Tsung-Dao Lee Institute, and School of Physics and Astronomy, Shanghai Jiao Tong University, Shanghai 200240, China}
\affiliation{${}^{3}$Department of Physics, National Taiwan University, Taipei 10617, Taiwan}
\affiliation{${}^{4}$Physics Division, National Center for Theoretical Sciences, Hsinchu 30013, Taiwan}

\begin{abstract}
The non-zero Dirac phases $\delta^q$  and $\delta^l$ in the CKM and PMNS mixing matrices signify CP violation. 
In general they are independent. Experimental data including recent T2K results show, however, that in the original KM parameterization for the mixing matrix, the sum $\delta^q_{\rm KM} + \delta^l_{\rm KM}$ is close to zero with $\delta^q_{\rm KM}$ to be approximately $\pi/2$. The KM parameterization may have provided some hints that these phases are actually related and CP is maximally violated. We show that this sum rule can be accommodated in models with spontaneous CP violation where both phases originate from a non-trivial common spontaneous CP violating maximized phase in the Higgs potential. We find some interesting phenomenological consequences for flavor changing neutral current and CP violation for such a model. 
In particular, data from $B_s - \bar B_s$ mixing provide very strong constraints on the mass scale for the new neutral scalars in the model, yet the model still allows the electric dipole moments of electron and neutron to reach to their current upper bounds. The model can be tested by near future experiments.
\end{abstract}

\maketitle

\noindent {\bf Introduction}

The CP violation has been observed in many experiments~\cite{Tanabashi:2018oca}. It is one of the crucial elements in explaining why our universe is one with matter dominating over anti-matter. However, the origin of CP violation is still a mystery.  In the standard model (SM) CP violating source in quark sector is due to the phase $\delta^q$ in the Cabbibo- Kobayashi-Maskawa (CKM) mixing matrix $V_{\rm CKM}$~\cite{Cabibbo:1963yz,Kobayashi:1973fv}. In the lepton sector, CP violating source  is due to the phase $\delta^l$ in Pontecorvo-Maki-Nakagawa-Sakata (PMNS) mixing matrix $V_{\rm PMNS}$~\cite{Pontecorvo:1957cp,  Pontecorvo:1957qd,Maki:1962mu}.  The mixings in the quark and lepton sectors may or may not be related~\cite{Minakata:2004xt}. It would be interesting that some mechanisms can relate CP violating phases $\delta^{q,l}$ so that they are coming from the same source~\cite{Tanimoto:2015hqa}. Among many possible origins of CP violation, spontaneous CP violation by vacuum is one of the very appealing possibilities~\cite{TDLI}. We find that spontaneous CP violation model can also relate the phases in the quark and lepton sectors. In this work, we construct a realistic invisible axion model based on PQ symmetry~\cite{PQ} which leads to a sum rule of $ \delta^q_{\rm KM} + \delta^l_{\rm KM}= 0$ with $\delta^q_{\rm KM}  = \pi/2$ in the original KM parameterization from experimental data, and the axion and neutrino mass seesaw scales are linked to each other. We also study some interesting implications which can be tested by future experiments.

For three generations, $V_{\rm CKM}$ and $V_{\rm PMNS}$ are $3\times 3$ unitary matrices and each can be parameterized by 3 rotation angles, $\theta_{12}$, $\theta_{23}$ and $\theta_{13}$ in the convention used by the Particle Data Group (PDG)~\cite{keung,Tanabashi:2018oca}, and a Dirac CP violating phase $\delta^q$ and $\delta^l$ for quark and lepton mixing matrices respectively. For Dirac neutrinos, $V_{\rm PMNS}$ matrix is similar to $V_{\rm CKM}$ in form. 
For Majorana neutrinos, one needs to multiply a diagonal
matrix $P = \mbox{diag}(1, e^{i \alpha_1/2}, e^{i\alpha_2/2})$ on right of $V_{\rm PMNS}$.
The values of rotation angles and the phases in the quark and lepton sectors are parametrization convention dependent.

There are a lot of information about quark and lepton mixing parameters. Their values are usually given in the PDG parameterization, for example quark and lepton mixing from the recent UTfit and Nufit Collaborations~\cite{UTfit2018, Esteban:2018azc,Nufit}, respectively. Concerning CP violating phases $\delta_{\rm PDG}^q$ and $\delta_{\rm PDG}^l$, the best (3$\sigma$ ranges) are given by
$\delta_{\rm PDG}^q/\pi =0.3717$ $(0.3606 \sim  0.3828)$ for quark mixing,
 and $\delta_{\rm PDG}^l / \pi =-0.772$ $(-1.200 \sim  - 0.017)$ for lepton mixing with normal 
 hierarchy (NH) (and $\delta_{\rm PDG}^l / \pi =-0.433$ $( -0.861\sim  -0.067)$ for inverted hierarchy (IH)). We see the PDG data allow the possibility that $\delta_{\rm PDG}^l =-\pi/2$. Furthermore, recent results from T2K also enforce such a possibility with $\delta_{\rm T2K}^l / \pi =-0.60^{+0.22}_{-0.18}$ (NH) and $-0.44^{+0.15}_{-0.17} $ (IH)~\cite{Abe:2019vii}.
Data, however, do not show correlations of rotation angles and phases in the two sectors.

The specific values of the rotation angles and phases are parametrization convention dependent. Let us consider the situation in the original KM parameterization for quark mixing~\cite{Kobayashi:1973fv}
\begin{eqnarray}
V_i = \left (
\begin{array}{ccc}
c^i_1&\;\;-s^i_1c^i_3&\;\;-s^i_1s^i_3\\
s^i_1c^i_2&\;\;c^i_1c^i_2c^i_3-s^i_2s^i_3 e^{i\delta^i_{\rm KM}}&\;\;c^i_1c^i_2s^i_3+s^i_2c^i_3 e^{i\delta^i_{\rm KM}}\\
s^i_1s^i_2&\;\;c^i_1s^i_2c^i_3+c^i_2 s^i_3 e^{i\delta^i_{\rm KM}}&\;\;c^i_1s^i_2s^i_3-c^i_2c^i_3 e^{i\delta^i_{\rm KM}}
\end{array}
\right )\;, \label{KMM}
\end{eqnarray}
where $s_j = \sin\theta_j$ and $c_j = \cos\theta_j$. Note that $V_i$ can be written in the form $V_{i1} + e^{i \delta^i_{\rm KM}}V_{i2}$ with $V_{i1,i2}$ real.

Using the values obtained by UTfit and Nufit collaborations ~\cite{UTfit2018, Esteban:2018azc,Nufit}, and the T2K results for the neutrino CP phase~\cite{Abe:2019vii}, we have for quark mixing,
\begin{eqnarray}
{\rm KM}:~~&&s_1^q=0.2250\;, \;\;(3\sigma:0.2240, 0.2260)\;, \;\;
s_2^q =0.03863\;, \;\; (3\sigma:0.03751, 0.03974)\;,
\nonumber\\
&&s_3^q =0.01633\;, \;\;(3\sigma:0.01584, 0.01683)\;, \;\;
\delta_{\rm KM}^q/\pi =0.4950\;,\;\;(3\sigma:0.4780, 0.5120)\;,
\end{eqnarray}
and for lepton mixing, 
\begin{eqnarray}
{\rm KM-NH}:~~&&s_1^l =0.5705\;, \;\;(3\sigma: 0.5383, 0.6048)\;, \;\;s_2^l =0.7894\;, \;\;(3\sigma:0.4530, 0.9101)\;,
\nonumber \\
&&s_3^l=0.2622\;,  \;\;(3\sigma:0.2372, 0.2885)\;, \;\;\delta_{\rm KM}^l/\pi =-0.5757\;, \;\; (3\sigma:-1, -0.0094)\;,
\nonumber \\
{\rm KM-IH}:~~&&s_1^l =0.5706\;, \;\;(3\sigma: 0.5385, 0.6050)\;, \;\;s_2^l =0.7202\;, \;\;(3\sigma:0.4677,0.8882)\;,
\nonumber \\
&&s_3^l =0.2634\;,  \;\;(3\sigma:0.2383, 0.2897)\;,\;\;\delta^l_{\rm KM}/\pi =-0.4275\;, \;\;(3\sigma:-0.8063, -0.1004)\;.
\end{eqnarray}

We see that in the KM parameterization, the phases are closer in size compared with those in the PDG parameterization and different in sign and $\delta^q_{\rm KM}$ is very close to $\pi/2$.  
Note that the current data allow the intriguing possibility that, in the neutrino mixing, $\theta^l_{23}$ and $\delta^l_{\rm PDG}$ to be $\pi/4$ and $-\pi/2$ (or $3\pi/2$). This has generated extensive efforts to realize such special scenarios which give some guidance to model buildings~\cite{mu-tau}. It has been pointed out that, in fact parameterization with a rotation angle to be $\pi/4$ and the CP violating phase to be $-\pi/2$ is not unique to PDG parameterization. The KM parameterization with $\theta_2^l = \pi/4$ and $\delta^l_{\rm KM}=-\pi/2$ is actually equivalent  to that  in the PDG parameterization with $\theta^l_{23}$ and $\delta^l_{\rm PDG}$ to be $\pi/4$ and $-\pi/2$ (or $3\pi/2$)~\cite{Pan:2019qcc}.

It is interesting to note that a sum rule emerges for the CP violating phases in the KM parameterization, namely, $\delta^q_{\rm KM} + \delta^l_{\rm KM} = 0$ within error bars and the central value of $\delta^q_{\rm KM}$ is very close to $\pi/2$. This might be a hint as a possible relation between CP violating phases in quark and lepton mixing matrices and CP is violated maximally in both quark and lepton sectors. They are related. We find that spontaneous CP violation model can accommodate this sum rule.
\\

\noindent {\bf Model realization of $\delta^q_{\rm KM} + \delta^l_{\rm KM} =0$ sum rule}

We now show that the sum rule of $\delta^q_{\rm KM} + \delta^l_{\rm KM} = 0$ can be realized in a multi-Higgs model which can solve the strong CP problem by Peccei-Quinn (PQ) symmetry with spontaneous CP violation. In addition, we can also relate the invisible axion PQ symmetry scale to the see-saw scale for small neutrino mass. In this model, beside the usual SM 3 generations of fermions $Q_L: (3,2,1/6)$, $U_R: (3,1,2/3)$, $D_R: (3,1,-1/3)$, $L_L: (1,2,-1/2)$ and $E_R: (1,1,-1)$, we also introduce 3 right handed neutrinos $\nu_R: (1,1,0)$ to facilitate seesaw mechanism for neutrino masses. Here the numbers in the brackets indicate the SM gauge group $SU(3)_C\times SU(2)_L\times U(1)_Y$ quantum numbers.
It has been shown that in order to have spontaneous CP violation with PQ symmetry at least three Higgs
doublets transforming as  $(1,2,-1/2)$: $\phi_i = e^{i\theta_i}H_i = e^{i\theta_i}((v_i +R_i +i A_i)/\sqrt{2}, h^-_i)^T$ with $i = 1,2,3$ and one complex Higgs singlet $(1,1,0)$:  $\tilde S = e^{i\theta_s}S = e^{i\theta_s}(v_s + R_s + iA_s)/\sqrt{2}$ are needed~\cite{He:1988dm,Geng:1988ty}. We will assume that $v_s\gg v_{1,2,3}$ so that the axion is invisible and also the seesaw mechanism is in effective. With the Higgs multiplets given, it has been shown that it is possible to have spontaneous CP violation with only one independent phase $\delta_{sp} = \theta_1 - \theta_2$ in the Higgs potential. We will not go into details here for the Higgs potential analysis.

For our purpose, we assign the following PQ charges to the Higgs fields and the fermion fields,
$
Q_L : 0\;,\;\;U_R: +1\;,\;\;D_R: +1\;,\;\;L_L : 0\;,\;\;\nu_R: +1\;,\;\;E_R: +1\;,\;\;\phi_{1,2}: +1\;,\;\;\phi_u=\phi_3: -1\;,\;\;\tilde S: +2\;.$ 
With the above PQ charges for the particles, the Yukawa couplings are given by
\begin{eqnarray}
L_Y = -\bar Q_L Y_u  \phi_3 U_R - \bar Q_L (Y_{d1} \tilde \phi_1 + Y_{d2}\tilde \phi_2) D_R -\bar L_L Y_\nu  \phi_3 \nu_R - \bar L_L (Y_{e1} \tilde \phi_1 + Y_{e2}\tilde \phi_2) E_R - {1\over 2} \bar \nu_R^c Y_s  \tilde S^\dagger \nu_R + H.C.\;,
\end{eqnarray}
where $\tilde \phi_i = -i\sigma_2 \phi^*_i$.

Absorbing the phases $\theta_3$, $-\theta_s/2$,  $-(\theta_3+\theta_s/2)$, $-\theta_1$ and $-(\theta_1+\theta_3+\theta_s/2)$ into redefinitions of $U_R$, $\nu_R$, $L_L$, $D_R$ and $E_R$, respectively, and writing the fermion mass terms in the form: $L_m =  - \overline{ D}_LM_d D_R -\overline{ U}_L M_u U_R
- \overline{ E}_LM_e E_R  - \overline{ L}_L M_D \nu_R- {1\over 2} \overline{ \nu_R^c} M_R \nu_R\;,$
we have
\begin{eqnarray}
M_d = M_{d1} + M_{d2}e^{i\delta_{sp}}\;, \;M_e = M_{e1}  + M_{e2} e^{i\delta_{sp}}\;,\;M_{ai} = Y_{ai} {v_{i}\over \sqrt{2}}\;;\; 
M_{u} = Y_u {v_3 \over \sqrt{2}}\;,\; M_D = Y_\nu {v_3\over \sqrt{2}}\;,\;M_R = Y_ s {v_s\over \sqrt{2}}\;,
\end{eqnarray}
The light seesaw neutrino mass matrix is given by $M_\nu = - M_D M_R^{-1}M^T_D $. The seesaw and axion scales are both determined by the $v_s$.

Working in the basis where $M_u$ and $M_\nu$ are already diagonalized, the mass matrices for the down quark and charged lepton can be written as
\begin{eqnarray}
M_d = V^{d \dagger}_L \hat M_d V^d_R\;,\;\;M_e = V^{e \dagger}_L \hat M_e V^e_R\;,
\end{eqnarray}
where $\hat M_i$ are diagonal matrices whose entries are the eigen-masses. One can identify
\begin{eqnarray}
V_q  = V^{d\dagger}_L\;,\;\;V_l = V^e_L\;.
\end{eqnarray}

We now try to find solutions so that  $\delta^{q,l}_{\rm KM}$ is  to be uniquely related to $\delta_{sp}$. We find that there exist a class of solutions allowing such a link to be achieved, provided that $M_{di} = V_{qi}\hat M_{d} V_R^{d}= \tilde M_{di} V_R^{d}$  and $M_{ei} = V_{li}^{\dagger}\hat M_{e} V_R^{e}= \tilde M_{ei} V_R^{e}$ with $i=1,2$. With this type of solutions $V_R^{q,l}$ can be absorbed into down-quark and charged lepton fields. We then have
\begin{eqnarray}
V_q\hat M_d = ({\rm Re} V_q + i {\rm Im} V_q)\hat M_d = (\tilde M_{d1} + \tilde M_{d2}e^{i\delta_{sp}}) \;,\;\;V^\dagger_l\hat M_e  = ({\rm Re} V^\dagger_l + i{\rm Im} V^\dagger_l)\hat M_e = (\tilde M_{e1} + \tilde M_{e2}e^{i\delta_{sp}})\;.
\end{eqnarray}

The above allows us to identify: $\delta^q_{\rm KM} = \delta_{sp}\;,\;\;\delta^l_{\rm KM} = - \delta_{sp}$. We therefore have obtained the desired sum rule: $\delta^q_{\rm KM} + \delta^l_{\rm KM}=0$. Since $\delta_{sp}$ is a spontaneous CP violation phase in the Higgs potential, by requiring CP violation to be maximal, $\delta^q_{\rm KM}$ is forced to take the value of $\pi/2$. We will work with such a solution described above. Note that the values of the elements in $\tilde M_{(d,e)i}$ are not constrained.

We should comment that although to obtain the desired solutions for the CP violating phases we have to pick up some special solutions, the fact that there are solutions which can accommodate experimental data shown in the KM parameterization linking the phases in quark and lepton sectors makes it interesting to study related phenomenological consequences further.
\\

\noindent {\bf New Higgs mediated interactions}

There are additional Higgs bosons in the model which bring in new interactions. To obtain new Higgs interactions, it is convenient to work in the basis where un-physical Higgs fields have been removed and the axion $a$ identified.
The un-physical Higgs bosons are the Goldstone fields $h_w$ and $h_z$ ``eaten'' by $W$
and $Z$ bosons. The physical fields, $a_{1,2}$, $a$ and $H^0_i$ related to the original fields are given by~\cite{Chen:2007nx}
\begin{eqnarray}\label{mixing}
&&\left (\begin{array}{c}A_1\\A_2\\A_3\\A_s\end{array}\right ) =
\left ( \begin{array}{cccc} v_2/v_{12}&\;\;-v_1v_3 v_s/v_{12}N_a&\;\;v_1/v&\;\;v_1v_3^2/vN_a\\
-v_1/v_{12}&\;\;-v_2v_3 v_s/v_{12}N_a&\;\;v_2/v&\;\;v_2v_3^2/vN_a\\
0&\;\;v_{12}v_s/N_a&\;\;v_3/v&\;\;-v_{12}^2v_3/vN_a\\
0&\;\;v_{12} v_3/N_a&\;\;0&\;\;vv_s/N_a\end{array}\right
)\left ( \begin{array}{c}a_1\\a_2\\h_z\\a \end{array}\right )\;,
\nonumber\\
&&\left (\begin{array}{c}h^-_1\\h^-_2\\h^-_3\end{array}\right ) =
\left ( \begin{array}{ccc} v_2/v_{12}&\;\;v_1v_3/v v_{12}&\;\;v_1/v\\
-v_1/v_{12}&\;\;v_2v_3/v v_{12}&\;\;v_2/v\\
0&\;\;-v_{12}/v&\;\;v_3/v\end{array}\right )\left (
\begin{array}{c}H^-_1\\H^-_2\\h_w \end{array}\right )\;,
\end{eqnarray}
where $v^2 = v^2_1+v^2_2 +v^2_3$, $v^2_{12} = v^2_1 +v^2_2$, and $N_a^2 = (v_{12}^2v_3^2 +
v_s^2v^2 )$.  $a_{1,2}$ and
$H^-_{1,2}$ are the physical degrees of freedom for the Higgs fields. With the same rotation as that for the neutral pseudoscalars, the neutral scalar Higgs fields $(R_1,R_2,R_3,R_s)^T$ become $(H_1^0,H_2^0,H^0_3, H^0_4)^T$. Since the invisible axion scale $v_s$ is much larger than the electroweak scale, to a very good
approximation, $N_a \approx v v_s$.

Note that $H^0_i$, $a_i$ and $H^-_i$ are not yet the mass eigenstates. To find the mass eigenstates, one needs to further analyze the Higgs potential. They are approximately mass eigenstates if the mixings are small.  In this limit $H^0_3 = h$ is the SM-like Higgs boson. 
The interacting terms of neutral Higgs boson with fermions are 
\begin{eqnarray}
L_{\rm Y} = && -\bar U_L {M_u\over v}U_R \left[{v_{12} v v_s\over v_3 N_a} (H^0_2 + i a_2) + H^0_3 - {v^2_{12}\over N_a} (H^0_4 + i a)\right]
\nonumber\\
&&-\left(\bar D_L {M_d\over v} D_R+\bar E_L {M_e\over v} E_R\right)\left[{v_2 v\over v_1 v_{12}}
(H^0_1 - i a_1) - {v_3 v v_s\over v_{12} N_a} (H^0_2 - i a_2)+ H^0_3 + {v^2_3\over N_a} (H^0_4 - i a)\right]\;
\nonumber
\\
&&+\left(\bar D_LV^\dagger_{q2} V_{q2}{M_d\over v} D_R+ \bar E_L V_{l2} V^\dagger_{l2} {M_e\over v} E_R\right) {v v_{12}\over v_1v_2} (H^0_1-ia_1)+ H.C..\;
\label{yukawa}
\end{eqnarray}
where we have decomposed $V_i= V_{i1} + V_{i2}e^{i\delta^i_{\rm KM}}$. The values of $V_{i1}$ and $V_{i2}$ can be read off from eq.~(\ref{KMM}).

In the above, we have not displayed the Yukawa couplings involving $\nu_R$ which has some components of light neutrinos, but the couplings are small. Furthermore, in the large $v_s$ limit, the axion is invisible and also the seesaw mechanism works. The couplings of $a$ and $H^0_4$ to SM fermions are also small. Note that $H^0_1$ and $a_1$ can mediate flavor changing neutral current (FCNC) at tree level~\cite{Chen:2007nx,Grossman:1994jb}. We will use data to constrain the model parameters from FCNC interactions due to exchange of $H^0_1$ and $a_1$. Due to spontaneous CP violation, the Higgs potential will mix $H^0_i$ with $a_i$ which also has important implications for CP violation and will be studied.

For definitiveness of numerical analysis, considering $v_3$ gives mass to top quark, $v_{1,2}$ are related to down-type quark masses with the bottom quark having the mass compatible with the larger one, 
to make Yukawa couplings to be large but not to upset perturbative calculations, we assume that the largest Yukawa couplings are around 1. It is then natural to have $v_3\sim v$, $v_{12}/v_3\sim m_b/m_t$. We also assume $v_2\sim v_1$, which implies $v_2/(v_1 v_{12}) \sim (1/v) (m_t/m_b)$ and $v_{12}/(v_1v_2) \sim (2/v)(m_t/m_b)$.  If $v_1 \neq v_2$,  the constraints obtained will be different. We will comment on this situation at the end of the numerical analysis.
\\

\noindent {\bf FCNC constraints}

The $V_{q2,l2}$ obtained from eq.~(\ref{KMM}) lead to FCNC only between the second and third generations which can cause $B_s-\bar B_s$ mixing and $\tau \to \mu \mu \bar \mu$.  The FCNC interactions can lead to enhanced $\tau \to \mu \gamma$ at loop level which gives the most stringent constraint on the scalar scale using data from lepton sector. 

\begin{figure}
\centering
\includegraphics[width=10cm]{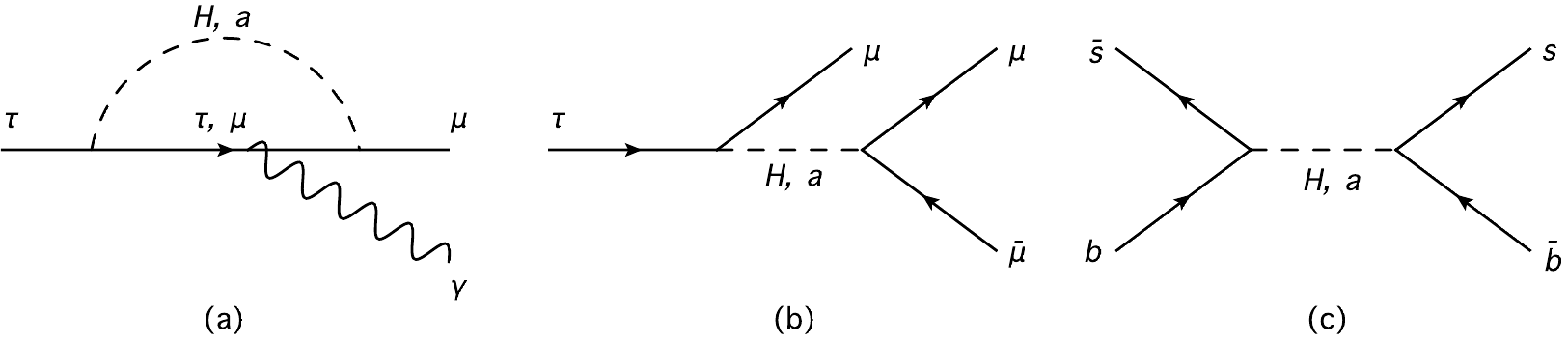}
\caption{The representative Feynman diagrams contributing to FCNC interactions.}
\label{fig1}
\end{figure}

The one loop diagram generating $\tau \to \mu \gamma $ is shown in Fig.~\ref{fig1}~(a). We find that the dominant contribution is from $\tau$ propagator due to the enhanced Yukawa couplings. Neglecting small corrections of order ${\cal O}(m_\mu^2/m_\tau^2)$, we have 
\begin{eqnarray}
\Gamma(\tau\to\mu\gamma)&=&{\alpha_{\rm em}m_\tau\over 64\pi^4}{(s^l_2c^l_2)^2\over 16}{m_\tau^4\over v_1^4} \left(1-(c^l_2)^2{v_{12}^2\over v_2^2}\right)^2\left[{m_\tau^2\over m_{a_1}^2}\left(\ln {m_\tau^2\over m_{a_1}^2}+ {5\over 3}\right)-{m_\tau^2\over m_{H_1}^2}\left(\ln {m_\tau^2\over m_{H_1}^2}+ {4\over 3}\right)\right]^2\;.
\label{taumuga}
\end{eqnarray}
where $\alpha_{\rm em}$ is the fine structure constant. 

\begin{figure}
\centering
\includegraphics[width=8.7cm]{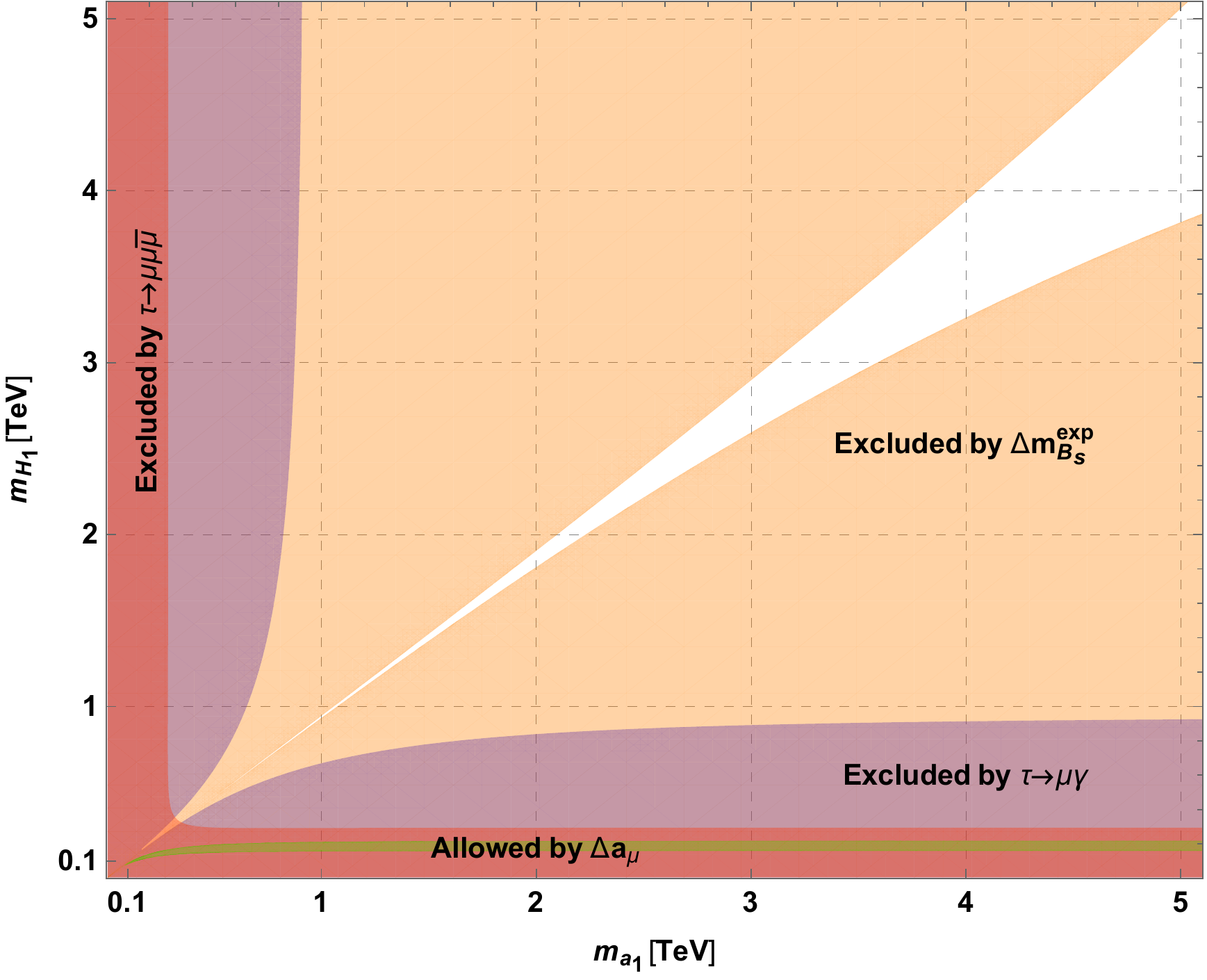}
\includegraphics[width=8.9cm]{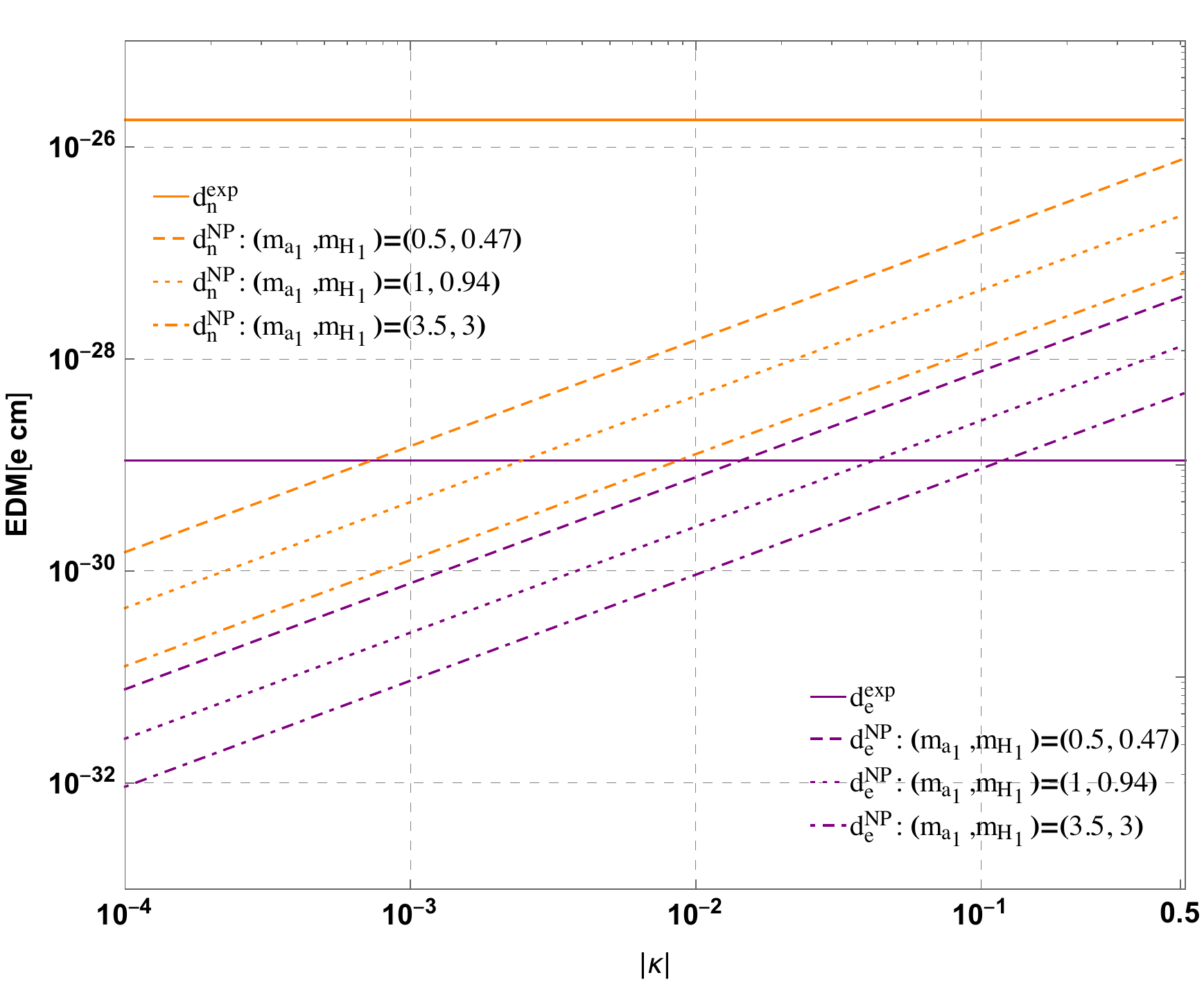}
\caption{Left panel: The excluded region in the $m_{a_1}-m_{H_1}$ plane. The excluded region by $\tau\to\mu\gamma$ is shown in purple, $\tau\to\mu\mu\bar\mu$ in red, $\Delta m_{B_s}$ in orange, respectively. The narrow green band is the allowed parameter space for explaining the muon $g-2$ anomaly. Right panel: The eEDM and nEDM as the function of mass mixing parameter $|\kappa|$.}
\label{fig2}
\end{figure}

Using the upper bound ${\rm Br}(\tau\to\mu\gamma)^{\rm exp}<4.4\times 10^{-8}$ at 90\% confidence level ({\rm CL})~\cite{Tanabashi:2018oca} with the central value of $s^l_2= 0.7894$ for the NH case, we obtain the excluded parameter space in the $m_{a_1}-m_{H_1}$ plane shown in the left panel of Fig.~\ref{fig2} in purple.  

For the $\tau \to \mu\mu \bar \mu$, it can take place at tree level as shown in Fig.~\ref{fig1} (b) via the exchange of $H_1^0$ and $a_1$. This process also receives comparable contributions from diagrams attaching the photon line in Fig.~\ref{fig1} (a) to a muon pair. Currently the experimental upper bound for the branching ratio is ${\rm Br}(\tau\to\mu\mu\bar\mu)^{\rm exp}<2.1\times 10^{-8}$ at 90\%~{\rm CL}~\cite{Tanabashi:2018oca}. Using this bound we have evaluated possible constraints on the masses of
$a_1$ and $H^0_1$ shown in Fig.~\ref{fig2} in red color. 

\begin{figure}
\centering
\includegraphics[width=10cm]{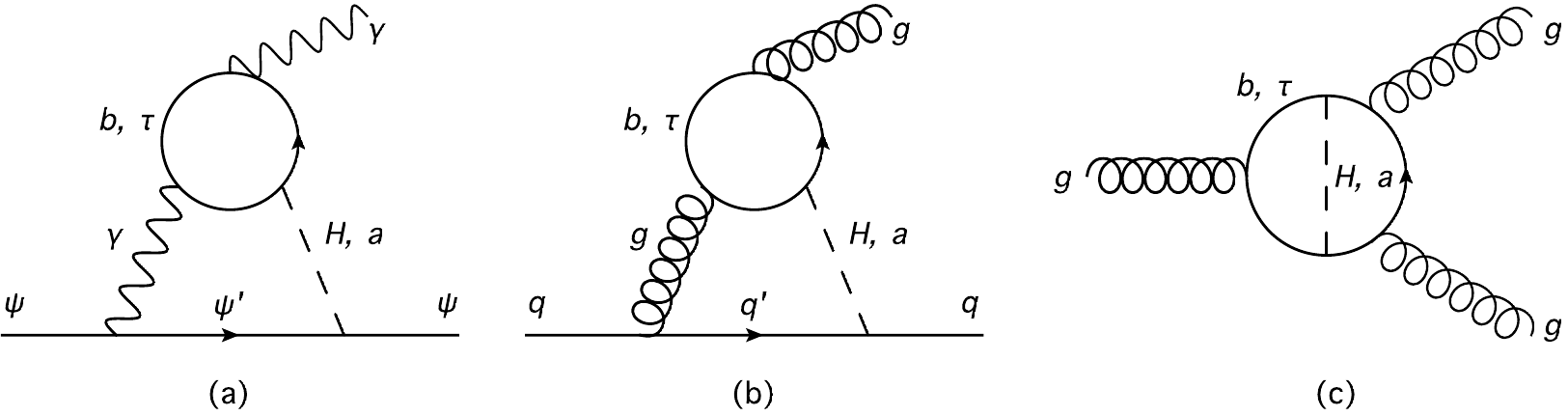}
\caption{(a). The Barr-Zee diagram contributing to the EDM of a fermion $\psi$. (b). The Barr-Zee diagram contributing to the cEDM of quark $q$. (c). The 2-loop mechanism generating Weinberg operator.}
\label{fig3}
\end{figure}

The enhanced coupling of $H_1^0$ and $a_1$ to leptons may also have impact on the anomalous magnetic dipole moment of muon $g-2$. We have the one and two loop contributions from Fig.~\ref{fig1} (a) with the initial tauon replaced by the muon and Fig.~\ref{fig3} (a) with the identification that $\psi=\mu$, respectively. We find the 1-loop contribution with an intermediate $\tau$ is dominant over the 2-loop contributions by a factor of ${\cal O}(10^3)$. As can be seen from Fig.~\ref{fig2}, with low mass of order 180~GeV, it is possible to produce correction $\Delta a_\mu \sim (28.02 \pm 7.37) \times 10^{-10}$~\cite{Keshavarzi:2019abf} to solve the muon g-2 anomaly problem. But this has been ruled out by other constraints. 

We find the mass difference $\Delta M_{B_s}$ of the $B_s - \bar B_s$ system provides the most stringent constraint. The SM has a well predicted value for $\Delta M_{B_s}$ with $\Delta M^{\rm SM}_{B_s} = (17.25\pm0.85)\;{\rm ps}^{-1}$~\cite{UTfit2018} which agrees with experimental data $\Delta M^{\rm exp}_{B_s} =(17.757\pm0.021)\;{\rm ps}^{-1}$~\cite{Tanabashi:2018oca} well. This  means that any new physics contributions are constrained. We will allow the new physics contribution $\Delta M^{\rm NP}_{B_s}$ and the SM prediction $\Delta M^{\rm SM}_{B_s}$ in the $3\sigma$ allowed ranges. Exchanges of $H_1^0$ and $a_1$  can contribute to $\Delta M^{\rm NP}_{B_s}$ at the tree level as shown in Fig.~\ref{fig1} (c). In the vacuum saturation approximation (VSA), we obtain $H_1^0$ and $a_1$ contributions to the mass splitting $\Delta M^{\rm NP}_{B_s}$ for $B_s -\bar B_s$ mixing as the following
\begin{eqnarray}
\Delta M^{\rm NP}_{B_s}&&=  {1\over 2}(s_3^qc_3^q)^2\left({vv_{12}\over v_1v_2}\right)^2\left\{ {5\over 12}\left( {1\over m_{H_1}^2}-{1\over m_{a_1}^2}\right)
{m_s^2+m_b^2\over v^2}{m_{B_s}^2 \over(m_s+m_b)^2} B_S\right .
\nonumber\\
&&-\left . \left({1\over m_{H_1}^2}+ {1\over m_{a_1}^2}\right){m_sm_b \over v^2}  \left[{m_{B_s}^2 \over (m_s+m_b)^2}B_S+{1\over 6} B_V \right]\right\}f_{B_s}^2m_{B_s}\;,
\end{eqnarray}
where $B_V$ and $B_S$ are the bag correction factors defined as via~\cite{Lenz:2006hd}:
$\langle \bar B_s| (\bar b\gamma_\mu P_L s)(\bar b\gamma^\mu P_L s)|B_s\rangle=(2 / 3)f_{B_s}^2m_{B_s}^2B_V$ and $
\langle \bar B_s| (\bar bP_R s)(\bar bP_R s)|B_s\rangle=-(5 /12){f_{B_s}^2m_{B_s}^4\over (m_s+m_b)^2}B_S$.
For numerical analysis, we take $B_V =0.849$ and $B_S =0.835$~\cite{DiLuzio:2019jyq} with $f_{B_s} =227.2\;{\rm MeV}$ for the $B_s$ decay constant, and $m_b =4.18\;{\rm GeV}$ and $m_s = 93\;{\rm MeV}$ for $b$ and $s$ quark masses. 

In the orange region of Fig.~\ref{fig2}, we display constraints from the above considerations. We find that the $B_s - \bar B_s$ mixing gives the most stringent constraint. The $H^0_1$ and $a_1$ masses are constrained to be larger than ${\cal O}(1~\rm TeV)$. This makes discovery of $H_1^0$ and $a_1$ at the LHC difficult. But some of the parameter space for the allowed masses for $H^0_1$ and $a_1$ may be probed by a 100 TeV collider.

If $v_1\neq v_2$, the constraint will be even stronger. As $\Delta M^{\rm NP}_{B_s}$ is proportional to $p=vv_{12} /(v_1 v_2)$, when $v_1$ becomes not equal to $v_2$ with a fixed $v_{12}$, the value $p$ will become larger and result in a stronger constraint on the masses for $H^0_1$ and $a_1$. Therefore the constraint provided above represents the most conservative one. 
\\

\noindent {\bf Electron and neutron EDM}

Due to CP violation in the Higgs potential, $H^0_{1,2,3}$ and $a_{1,2}$ mixing will be generated at tree level through terms in the potential such as $(H^\dagger_1 H_2)^2 e^{-i 2\delta_{sp}}$, $(H^\dagger_1 H_1) (H_1^\dagger H_2)e^{-i \delta_{sp}}$, $(H^\dagger_2 H_2) (H_1^\dagger H_2)e^{-i \delta_{sp}}$, $(H^\dagger_3 H_3) (H_1^\dagger H_2)e^{-i \delta_{sp}}$ and $(H^\dagger_3 H_2) (H_1^\dagger H_3)e^{-i \delta_{sp}}$. 
We will parameterize the mixing approximately by $H^0_i \to H^0_i + \kappa_{ij} a_j$ and $a_i \to a_i- \kappa_{ji} H^0_j$. The mixing parameters $\kappa_{ij}$ are free parameters which depend on the parameters in the potential. If the mixing is mainly due to $a_i$ and $H_j$ mass mixing term $m^2_{ij}$,  $\kappa_{ij}$ is given approximately by  $m^2_{ij}/(m^2_{a_j} - m^2_{H_i})$ which we assume to be much smaller than 1.  An interesting effect of such a mixing is that a non-zero electric dipole moment (EDM) $d_f$ of a fermion $f$ will be induced at loop levels. 

For the electron EDM (eEDM), the 1-loop contribution is similar to Fig.~\ref{fig1} (a) with the external $\tau$ and $\mu$ being substituted by the electron. However, the dominant contribution is from the 2-loop Barr-Zee type diagram~\cite{Barr:1990vd} shown in Fig.~\ref{fig3} (a) with the $b$-quark and $\tau$ lepton circulating in the loop, we have
\begin{eqnarray}
d_e^{2L}= {e\alpha_{\rm em}\over 96\pi^3}m_e\left[G(m_b,e)+3G(m_\tau,e)\right]\;,
\end{eqnarray}
where
\begin{eqnarray}
G(m,e)&&=2{\kappa v_2^2\over v_1^2v_{12}^2}\left(1-c_m^2{v_{12}^2\over v_2^2}\right)\left[f\left({m^2\over m_{a_1}^2}\right) +g\left({m^2\over m_{a_1}^2}\right)-f\left({m^2\over m_{H_1}^2}\right)-g\left({m^2\over m_{H_1}^2}\right) \right]\;,
\end{eqnarray}
with $c_m=c_3^q (c_2^l)$ for $m=m_b (m_\tau)$. The loop function $f(z)$ and $g(z)$ can be found in~\cite{Chen:2007nx}.

For the neutron EDM (nEDM), there are several contributions such as those from (a) the EDM $-{i \over 2}d_q \bar q\sigma_{\mu\nu}\gamma_5 qF^{\mu\nu}$, (b) the color EDM (cEDM) $-{i \over 2}f_q g_s\bar qT^A\sigma_{\mu\nu}\gamma_5 qG^{A,\mu\nu}$, and (c) the Weinberg operator ${1\over 3}C_Wf^{abc}G_{\mu\nu}^a\tilde G^{b,\nu\beta} G_\beta^{c,\mu}$~\cite{Weinberg:1989dx}.
The nEDM from the above effective interactions are estimated to be~\cite{Chen:2007nx}
\begin{eqnarray}
d_{n}\approx\eta_d\left({4\over 3}d_d-{1\over 3}d_u\right)_\Lambda+e\eta_f\left({4\over 9}f_d+{2\over 9}f_u\right)_\Lambda+ef_\pi \xi C_W\;,
\end{eqnarray}
where $f_\pi=95~\rm MeV$ is the pion decay constant. The QCD running factors from the electroweak to the hadronic scale $\Lambda\sim1~\rm GeV$ are approximately $\eta_d\approx 0.166, \eta_f\approx 0.0117$, and $\xi\approx1.2\times10^{-4}$, respectively.

In our model, the dominant contribution is from 2-loop diagrams shown in Fig.~\ref{fig3} which contribute the EDM in (a), the cEDM in (b), and the Weinberg operator in (c), respectively.
The 2-loop contributions to EDM and cEDM of up quark from the mixing of $H_1^0$ and $a_1$ vanish, i.e., $d_u=f_u=0$. Then we have the following compatible contributions
\begin{eqnarray}
d_d^{2L}&=&{e\alpha_{\rm em}\over 288\pi^3}m_d\left[G(m_b,d)+3G(m_\tau,d)\right]\;,\;\;
f_d^{2L}=-{\alpha_s\over 64\pi^3}m_dG(m_b,d)\;,\nonumber\\
C_W^{2L}&=&-{1 \over 4\pi}{\kappa v_2^2\over v_1^2v_{12}^2}\left(1-(c_3^q)^2{v_{12}^2\over v_2^2}\right)^2\left[ h\left({m_b^2\over m_{a_1}^2}\right)-h\left({m_b^2\over m_{H_1}^2}\right)\right].
\end{eqnarray}
The loop function $h(z)$ can be also found in~\cite{Chen:2007nx}.

In the right panel of Fig.~\ref{fig2}, we show the eEDM in purple and nEDM in orange as the function of the CP violating parameter $\kappa$. The solid lines represent the current experimental limits, in which $d_e^{\rm exp}<1.1\times 10^{-29} e\;{\rm cm}$~\cite{Andreev:2018ayy} and $d_n^{\rm exp}<1.8\times 10^{-26} e\;{\rm cm}$~\cite{Abel:2020gbr} at 90\% CL. The other different types of lines represent the different choices of $m_{a_1}$ and $m_{H_1}$ allowed by the $B_s - \bar B_s$ mixing constraint.  One can see, when $|\kappa|$ runs from $10^{-4}$ to $0.5$, the eEDM and nEDM both get improved by several orders of magnitude relative to the SM predictions where $d_e^{\rm SM}\leq{\cal O}(10^{-39})\;e\;{\rm cm}$~\cite{Pospelov:2005pr,Yamaguchi:2020eub} and $d_n^{\rm SM}\sim{\cal O}(10^{-32})\;e\;{\rm cm}$~\cite{nEDM}, which could be reached by future EDM experiments to test such possibilities.


\section*{Acknowledgments}
This work was supported in part by Key Laboratory for Particle Physics, Astrophysics and Cosmology, Ministry of Education, and Shanghai Key Laboratory for Particle Physics and Cosmology (Grant No. 15DZ2272100), and in part by the NSFC (Grant Nos. 11575111 and 11735010), and in part by the MOST (Grant No. MOST 106-2112-M-002-003-MY3 ). This work was also supported in part by the Australian Government through the Australian Research Council.

\end{document}